\documentstyle[12pt,preprint,aps]{revtex}
\begin{document}

\title{Light Meson Mass Spectrum from Gauge Fields on Supergravity Backgrounds
}
\author{M. A. Hartman and 
C.~C.~Taylor}
\address{Department of Physics, Case Western Reserve University,  Cleveland,
Ohio 44106-7079}

\date{\today}


\maketitle
\begin{abstract}

We propose that the spectrum of light mesons (the $\pi$ and $\rho$, together
with their radial excitations) can be calculated in the limit of
vanishing light quark masses by studying gauge theory
(open string theory) on 
suitable higher dimensional background geometries.  Using the 
metric proposed by Witten for glueball calculations as a paradigmatic example,
we find a spectrum which is in 
startlingly good agreement with the masses tabulated by the Particle Data Group 
\cite{PDG}.  These calculations have only one free parameter, corresponding
to the overall mass scale.
We make predictions for the next several particles in the spectrum.
\end{abstract}
\pacs{11.25.Sq,12.38.-t,12.38.Qk,
14.40.-n}


String theory arose as an attempt to understand the strong interactions.
With the advent of gauge theories in the early 1970's, string theory 
metamorphosed into a ``theory of everything", and, until recently, 
has had little new to say about strong interaction phenomenology.  

The last two years, however, have seen dramatic progress towards understanding
the relationship between string theory and gauge theories. Maldacena's conjecture
of the relationship between ${\cal N}=4$ supersymmetric
gauge theories in four dimensions and string theory
(supergravity) on $AdS^5 \times S^5$ has proven very fruitful
\cite{Maldacena,Witten,Gubser:1998bc,review}.  Witten's 
conjectured extension to
non-supersymmetric theories \cite{Witten2}
has provided the basis for calculations of glueball
mass spectra which have been directly compared to the results of
lattice calculations with
good numerical agreement \cite{Csaki:1998qr,deMelloKoch:1998qs,Zyskin:1998tg}.

Unfortunately, the calculations presented to date correspond to 
theories which do not contain states with
light quarks, and so are not directly relevant to the real world.  Indeed,
the question as to how to create such models starting with string theory
or M-theory is still unclear.  

There is, however, a qualitative line of reasoning to pursue.  QCD flux
tubes are presumably the starting point for a stringy picture of the strong
interactions.  The extra dimensions involved in the Maldacena/Witten conjectures
plausibly correspond in some fashion to internal degrees of freedom (such as thickness)
of the flux tube.  It is thus reasonable that masses of glueballs, which are
closed flux tubes in a QCD picture, can be calculated in terms of closed 
(super)string
modes on non-trivial higher dimensional backgrounds, such as those proposed
by Witten and others.  

Mesons, on the other hand, should be associated with
the modes of open strings.  This correspondance  was made in the early days
of string theory, and is also suggested by a constituent quark
picture in which the quark and
antiquark of a meson are connected by a QCD flux tube.  Since the ends of
open strings move at the speed of light, this picture suggests that we will
be modeling states with massless quarks; i.e. chiral symmetry is spontaneously,
but not explicitly, broken.
If the dynamics of these meson flux tubes are locally
the same as those for glueballs, then meson dynamics in the real world
should correspond to open string dynamics on the same spaces considered in the
glueball calculations.

The purpose of this 
paper is to explore the implications of this suggestion. We do this by 
studying the mass spectrum associated with the lowest mode of the open
(super)string:   a massless gauge field $A_\mu$.  Our calculations 
closely follow the calculations of the glueball mass, so our description will
be brief.

We consider the action
\begin{equation}
I = -{1\over 4} \int d^{10} x \sqrt{g} F_{\mu \nu} F^{\mu \nu} \label{action}
\end{equation}
in Witten's black hole metric \cite{Witten2}
\begin{equation}
{ds^2\over {l_s^2 g_5^2 N/4 \pi}} = {dr^2\over r (1 - {1\over r^6})}
+ r^3(1 -{1\over r^6})d\tau^2 +r^3 \sum_{i=1}^4 dx_i^2 +r d\Omega_4^2.
\label{metric}
\end{equation}
Though the interacting theory will be expressed in terms of a non-abelian field
strength, we only need the quadratic part of the Lagrangian to calculate
masses and so we take
$F_{\mu \nu} = \partial_{\mu} A_\nu - \partial_\nu A_\mu$.  We similarly suppress
the isospin indices of the $\pi$ and $\rho$ fields, and the corresponding indices
on the gauge field $A_\mu$.

We choose a gauge in which $A_r = 0$, and make an ansatz for the modes of 
interest:
\begin{eqnarray}
A_\tau&= &H(r) e^{ikx}, \\
A_i &= &\epsilon_i(k) G(r) e^{ikx}, \label{ansatz}
\end{eqnarray}
where $\epsilon_i$ defines the polarization of the vector mesons.
The equations of motion then imply that $\epsilon \cdot k=0$ and that
\begin{eqnarray}
(r^6 -1)\partial_{r}[r^7 \partial_{r} H(r)] - k^2 r^9 H(r)=0,
\label{pieqn}\\
\partial_{r} [ (r^7 -r) \partial_{r} G(r)] - k^2 r^3 G(r) =0.
\label{rhoeqn}
\end{eqnarray}

The physical $\pi$ and $\rho$,
together with their radial excitations, thus correspond respectively to solutions of
the equations (\ref{pieqn})  and (\ref{rhoeqn})
which are regular at $r=1$ (the horizon of the
black hole), and which are normalizable, i.e. fall off sufficiently rapidly as
$r \to \infty$.  The masses are given by the corresponding eigenvalues,
$m_i^2 = -k_i^2$.

An interesting feature of eqn. (\ref{pieqn}), in
contrast to eqn. (\ref{rhoeqn}) and the corresponding glueball equations,
is that it has
an acceptable solution for $k^2=0$ given by $H(r) = A/r^6$.  There
is thus a massless pion, as expected from the above general
considerations.  

Other particle masses are found by numerical techniques.  We have calculated
the series solution about $r=1$ (a regular singular point of our equations),
and used this to determine the functions and their derivatives at $r=1.01$.
These values are then used as input to numerically integrate the equations.
Since normalizable solutions vanish for large $r$, while non-normalizable ones
do not, 
an eigenvalue is signaled by a change in sign of the solution at large $r$ as 
$k^2$ is changed.  Successive eigenvalues should correspond to successive excitations
of the $\pi$ (resp. $\rho$).  We have tested this technique by calculating
the glueball spectrum, and reproduce
the precise results of \cite{Zyskin:1998tg} to the accuracy quoted here.

The overall mass scale is not determined by our equations.   We fix this by
comparison with the data, minimizing $ \sum_i ((m_i^{obs})^2 -\lambda
(m^{calc}_i)^2)^2$ with respect to $\lambda$.  The result, $\lambda=.043$,
is not very different if, for example,
 one minimizes the differences between masses rather
than squared-masses.  

In Table \ref{compare} we tabulate our results
together with the observed masses of the corresponding particles.  
The resulting agreement is surprisingly good.  The worst agreement is for
the $\pi(135)$ and $\rho(770)$, and is presumably a consequence of the fact that 
there is no explicit chiral symmetry breaking in this model. 
The general pattern is consistent with
quark-model expectations: the mass of the $\rho$ will be high if
the mass of the $\pi$ is low, and these effects will be less important at
higher masses. (This, incidentally, is why normalizing to the mass of the
$\rho$ is probably not the best way of comparing the calculations to the
data.)

It is worth while making a few comments about the states that we have included.
The eigenvalues of the equations (\ref{pieqn}) and (\ref{rhoeqn}) correspond,
respectively, to masses of successive radial excitations of the $\pi$ and $\rho$.
The $\rho(1700)$, which has the same overall quantum numbers as the $\rho$,
is a D-wave state \cite{PDG}, 
and thus an orbital excitation of the $\rho$ which should
not be reproduced by the present calculation.  The
$\rho(2150)$, which we have included, is presented in the full Meson Particle
Listings  but is omitted in the Data Summary
Table of the Review of Particle Physics.  
Since we did not include this mode
in determining the scale, it is in some sense a prediction of the present
calculations.  

There are no other states in the Meson Particle Listings with
quantum numbers such that they should be reproduced by this calculation.
We do, however, predict additional radial excitations
of the $\pi$ at  2.32 GeV and 2.88 GeV and of the $\rho$ at 2.64 GeV.   
Though it may be difficult to isolate
such states, it would be extremely interesting to confirm (or deny)
their existence.  An infinite
number of additional states are also predicted, but all have masses larger
than 3 GeV.

We believe that these results strongly suggest that 
present string theory technology is directly relevant to
strong interaction phenomenology.  A number of directions for further work
immediately suggest themselves.  

First, we have not discussed the Kaluza-Klein
modes associated with the $S^4$ directions.  In the glueball case, modifications
of the metric (\ref{metric}) corresponding to rotating D-branes decouple the
unwanted Kaluza-Klein modes without significantly changing the masses of the
physical glueballs \cite{Csaki:1998cb}.  We expect similar results here, 
but this must be checked.

Second, the fact that the pion is massless in this model is extremely interesting.
In addition to studying the implications of this, 
one should also study the impact of adding an explicit mass
term to the action (\ref{action}).  Generalizations of this may permit the
extension of this model to $K$ and $K^*$ mesons, and perhaps others.

Beyond this, it may be possible to study interactions in this model. Since baryons
can already be understood as topological excitations of pions in Skyrme models
\cite{Adkins:1983ya},
perhaps a similar understanding is possible here.  Finally, other modes of the
open string should be studied; they should correspond to other mesons, including
orbital excitations of the $\pi$ and $\rho$.

We hope to report on some of these issues in the near future.

We would like to thank Mark Trodden for helpful conversations, and
Dan Akerib, Zlatko Dimcovic, Glenn Starkman and Tanmay Vachaspati for
readings of the manuscript.  We would also like to thank Juan Maldacena, Jeffrey
Harvey and Antal Jevicki for useful background discussions.
This work was supported in part by the National Science Foundation 
through grant PHY-9940415.


%
%

%
%

\begin{table}[h]
\caption{Light meson masses, as calculated and as reported by the 
Particle Data Group [1]. The calculated mass of the pion is an analytical
result.  The other masses are calculated by the shooting method.
A single overall scale factor is used to fit to the data. Note that
the $\rho(2150)$ is omitted from the Data Summary Table of the Review
of Particle Physics and is not used here in adjusting the scale parameter.}
\label{compare}
\begin{tabular}{|l|ll|}
particle 	& calculated mass (GeV) 	& observed mass (GeV) \\\hline
$\pi$		& $0 	$			& $0.135$		\\
$\rho(770)$	& $0.97	$			& $0.770$		\\
$\pi(1300)$	& $1.17	$			& $1.300 \pm 0.100$		\\
$\rho(1450)$& $1.54	$			& $1.465 \pm 0.025$		\\
$\pi(1800) $& $1.76 $			& $1.801 \pm 0.013$		\\
$\rho(2150)$& $2.10 $			& $2.149 \pm 0.017$		\\
$\pi(?)$	& $2.32 $			& ?						\\
$\rho(?)$	& $2.64 $			& ?						\\
$\pi(?)	$	& $2.88 $			& ?						\\
\end{tabular}
\end{table}

\end{document}